\documentclass[12pt]{article}
\usepackage{fullpage}
\usepackage{setspace}
\usepackage{lscape}
\usepackage{amsmath}
\usepackage{amsthm}
\usepackage{acronym}
\usepackage{graphicx}
\usepackage{amssymb}
\usepackage{bm}
\usepackage{mathtools}
\usepackage{booktabs}
\usepackage{multirow}
\usepackage{hyperref}
\usepackage{setspace}
\usepackage{natbib}
\usepackage{color}
\usepackage{caption}
\usepackage{amsfonts}
\usepackage{bm}
\usepackage{mathtools}
\usepackage{booktabs}
\usepackage{mathrsfs}
\usepackage{mathptmx}
\usepackage{multirow}
\usepackage{mathrsfs}
\usepackage{subcaption}
\usepackage{latexsym}
\usepackage{nicefrac}
\usepackage{verbatim}
\usepackage{lineno}

\usepackage{hyperref}
\let\svthefootnote\thefootnote
\newcommand\freefootnote[1]{%
  \let\thefootnote\relax%
  \footnotetext{#1}%
  \let\thefootnote\svthefootnote%
}

\everydisplay{\abovedisplayshortskip\belowdisplayshortskip}
\makeatletter \oddsidemargin  -.1in \evensidemargin -.1in

\renewcommand{\thefootnote}{\fnsymbol{footnote}}

\theoremstyle{plain}
\newtheorem*{Lemma*}{Lemma}
\title{Bayesian Analysis of  the Beta Regression Model Subject to Linear Inequality Restrictions with Application}
\author{Solmaz Seifollahi$^{1}$ \footnote{Email: s.seifollahi@tabrizu.ac.ir (Corresponding author).},  
Hossein Bevrani$^{1}$ \footnote{Email: bevrani@tabrizu.ac.ir.}, Kristofer M{\aa}nsson $^{2}$ \footnote{Email: kristofer.mansson@ju.se.}\\
\small {$^{1}$ Department of Statistics, University of Tabriz, Tabriz, Iran}\\
\small {Department of Economics, Finance and Statistics, J\"{o}nk\"{o}ping International Business School, J\"{o}nk\"{o}ping}
}

\date{}

\begin{document}
\maketitle
\noindent{\textbf{{\em Abstract:}}}
Recent studies in machine learning are based on models in which parameters or state variables are bounded restricted. These restrictions are from prior information to ensure the validity of scientific theories or structural consistency based on physical phenomena.
The valuable information contained in the restrictions must be considered during the estimation process to improve estimation accuracy.
Many researchers have focused on linear regression models subject to linear inequality restrictions, but generalized linear models have received little attention. In this paper, the parameters of beta Bayesian regression models subjected to linear inequality restrictions are estimated. The proposed Bayesian restricted estimator, which is demonstrated by simulated studies, outperforms ordinary estimators. Even in the presence of multicollinearity, it outperforms the ridge estimator in terms of the standard deviation and the mean squared error. The results confirm that the proposed Bayesian restricted estimator makes sparsity in parameter estimating without using the regularization penalty. Finally, a real data set is analyzed by the new proposed Bayesian estimation method.
\vskip 3mm
\noindent {\bf {\em  Keywords and phrases:}} Bayesian inference, Beta regression model, Linear inequality restrictions,  Link function, Restricted estimator.\\
\noindent{{\bf{\em MSC2020 subject classifications:}} 62F15; 62J12.}

\section{Introduction}
Beta regression models (BRMs), proposed by \cite{Ferrari1}, have become popular to model response variable which is bounded over $(0, 1)$.
Various applications of BRMs have been studied for instance the percentages of body fat (\cite{Erkoc}),  the proportion of crude oil after distillation and fractionation (\cite{Qasim}), the color characteristics of hazelnuts (\cite{Karlsson}).

Commonly, the maximum likelihood estimator is used to estimate model parameters. However, in certain applications, it is beneficial to incorporate prior information about the parameters in the model. The inclusion of such information can enhance the accuracy of the estimation procedure.
The prior information can be defined as restrictions in models, which may take the form of linear or nonlinear constraints, as well as equality or inequality.

In BRMs, linear equality restrictions have been investigated by \cite{Seifollahi} and \cite{Arabi}. They aimed to enhance the Beta Maximum Likelihood Estimator (BMLE) and Beta Liu Estimator (BLE), respectively, by employing shrinkage methods such as James-Stein, positive James-Stein, and preliminary test methods.
Compared to linear equality restrictions, there are situations where it becomes necessary to impose linear inequality constraints on regression parameters. These constraints ensure structural consistency based on physical phenomena or the validity of scientific theories.
In applied econometrics, for instance, there are cases where certain coefficient parameters must be non-negative or non-positive (\cite{Pindyck}; \cite{Bails}). Due to physical considerations in hyper-spectral imaging (\cite{Manolakis}), the coefficient parameters should be non-negative. \cite{Wang} provide another example of applications of these restrictions in astronomy and zoology, and \cite{Zhu} in geodesy.

In the classical framework, the linear regression models subjected to the linear inequality restrictions have been widely investigated (see
\cite{Judge}, \cite{Lovell}, \cite{Escobar1, Escobar2}, and \cite{Ohtani}). Recently, in the era of big data, it has been proved that the incorporation of non-negative restrictions provides sparsity without using any regularization in linear regression models (\cite{Meinshausen}, and \cite{Slawski}) and also in generalized linear models (GLMs) (\cite{Koike}).
On the other hand, Bayesian models provide a straightforward approach to incorporating linear inequality restrictions in the estimation process. Several studies have investigated Bayesian inferences on linear regression models subject to linear inequality restrictions. Some notable works in this area include \cite{Geweke1, Geweke2}, \cite{Davis}, \cite{Neelon} and recently in \cite{Veiga} and \cite{Seifollahi2}.
Most of these studies are about the multiple linear regression models however, such restrictions  can occur in applications where GLMs are applicable. 
\cite{Ghosal} had introduced an algorithm to obtain the Bayesian estimation based on the linear inequality restrictions in the GLMs but it relies on certain conditions that may not always be satisfied. For instance, these conditions are not met in the case of BRMs with a logit link function. However, \cite{Seifollahi3} tried to introduced an algorithm in GLMs to take into account the linear inequality restrictions by using any link function in Gamma regression models.
Therefore, there is a need for a practical method that allows for Bayesian inference in BRMs subject to linear inequality restrictions.
The aim of this paper is precisely to address this gap and focus on Bayesian inference in BRMs subjected to linear inequality restrictions.

The structure of the paper is as follows:
The BRMs and the maximum likelihood estimator of the regression parameters are presented in section \ref{sec2}.
Section \ref{sec3} introduces our Bayesian estimation method for BRMs subjected to linear inequality constraints.
In Section \ref{sec4}, we compare the performance of our proposed Bayesian estimator to existing methods using two simulation studies.
Section \ref{sec5} contains an analysis of a real-life data set. Finally, in Section \ref{sec6}, we conclude the paper.

\section{Beta Regression: Model and Estimation}\label{sec2}
Beta regression model for a response variable confined to $(0, 1)$ was first proposed by \cite{Ferrari1} who used a monotone differentiable function called the link function to connect the mean of the response variable to a set of independent variables.
 Assume $y_i$ is a continuous random variable following the Beta probability density function:
\begin{equation}\label{md1}
  f(y_i)= \dfrac{\Gamma(\gamma)}{\Gamma(\mu_i \gamma) \Gamma\left((1-\mu_i) \gamma\right)} y_i^{\mu_i \gamma-1} (1-y_i)^{(1-\mu_i) \gamma-1}; \qquad i:1, 2, \cdots, n,
\end{equation}
where $0<y_i, \mu_i<1$ and $\gamma>0$. $\Gamma(.)$ is a gamma function and $\gamma$ is called the precision parameter which is supposed to be constant and known through observations in this paper.
The model makes it possible for the response mean to depend on linear predictors in the following way by applying the link function $g(.)$ :
\begin{equation*}
  g(\mu_i)= \log(\dfrac{\mu_i}{1-\mu_i})= X_i^T \pmb{\beta}= \eta_i; \qquad i:1, 2, \cdots, n
\end{equation*}
where $\pmb{\beta}=(\beta_1, \beta_2, \cdots, \beta_p)^T$ is a vector of $p$ unknown parameters and $X_i=(x_{i1}, x_{i2}, \cdots, x_{ip})^T$ is the $i$th observations for the covariates and $\eta_i$ is called linear predictor.
To find the estimation of $\pmb{\beta}$, the log-likelihood function is required. For BRM, the log-likelihood function is defined as:
\begin{align}\label{log:lik}
 \ell (y|\pmb{\beta}, \gamma, X) &= n\log \big(\Gamma(\gamma)\big) +  \sum_{i=1}^{n}  \bigg\{ (\gamma-1)  \log(1-y_i)- \log \log(y_i)\bigg\}\nonumber\\
                     & \qquad  + \sum_{i=1}^{n} \bigg\{\gamma\mu_i  \log(\dfrac{y_i}{1-y_i})- \log \big(\Gamma(\gamma \mu_i)\big)- \log \big(\Gamma(\gamma (1-\mu_i))\big)\bigg\}
\end{align}
The Beta maximum likelihood estimator (BMLE) of vector $\pmb{\beta}$ is obtained by the iterative re-weighted least square (IWLS) method as
\begin{equation}\label{MLE}
\hat{\pmb{\beta}}_{BMLE}= (X^T \hat{\pmb{C}} X)^T X^T \hat{\pmb{C}} \pmb{U},
\end{equation}
where
\begin{align}
\hat{\pmb{C}}&= diag(C_1, C_2, \cdots, C_n),\\
C_i & = \gamma \bigg\{ \Psi' (\hat{\mu}_i \gamma)+ \Psi' ((1-\hat{\mu}_i) \gamma) \bigg\} \dfrac{1}{\big\{g'(\hat{\mu}_i)\big\}^2},\\
\pmb{U}&= \hat{\pmb{\eta}}+ \hat{\pmb{C}}^{-1} \hat{T}(\tilde{\pmb{y}}- \tilde{\pmb{\mu}}),\\
\hat{T}&=diag(\dfrac{1}{g'(\hat{\mu}_1)}, \dfrac{1}{g'(\hat{\mu}_2)}, \cdots, \dfrac{1}{g'(\hat{\mu}_n)}),\\
\tilde{\pmb{\mu}}&= (\tilde{\mu}_1, \tilde{\mu}_2, \cdots, \tilde{\mu}_n)^T; \qquad \tilde{\mu}_i= \Psi(\hat{\mu}_i \gamma)- \Psi((1-\hat{\mu}_i)\gamma),\\
\tilde{\pmb{y}}&= (\tilde{y}_1, \tilde{y}_2, \cdots, \tilde{y}_n)^T; \qquad \tilde{y}_i=  \log(\dfrac{y_i}{1-y_i}),
\end{align}
and $\Psi(.)$ denotes digamma function.

\section{Bayesian Inference in Restricted Beta Regression}\label{sec3}
First, we commence with the model such that $ X^T \hat{\pmb{C}} X$ is non-singular. Then, in the next section, by using the simulation study, we illustrate that when $ X^T \hat{\pmb{C}} X$ is singular, the Beta Bayesian estimator based on the linear inequality restrictions acts as a natural penalty in the estimation process.
 The considered restrictions on the model parameters are bounded as follows:
\begin{equation}\label{H0}
  \pmb{H}\pmb{\beta} \leq \pmb{G},
\end{equation}
where $\pmb{H}$ and $\pmb{G}$ are respectively a pre-specified $q \times p $ matrix and vector of $q$ length. Here, the value of $q$, where $q$ is the number of restrictions, is allowed to be more than $p$. It is supposed that the sub-space created by restrictions in \eqref{H0} is not empty.
In traditional Bayesian inference, without any restrictions on the parameters of the model, the multivariate normal distribution is chosen as the prior distribution of the vector of parameters, $\pmb{\beta}$, such as
\begin{equation}\label{MVN}
  \pmb{\beta} \sim N_p (\pmb{\mu}_{\pmb{\beta}}, \pmb{\Sigma}_{\pmb{\beta}}),
\end{equation}
but here, there is a strongly belief that the model parameters satisfy in \eqref{H0}. Therefore if \eqref{MVN} is used as a prior distribution of the parameters, the unrestricted Bayesian estimator will be lost efficiency. To make it clear, suppose that $Y_i \sim N (\mu , \sigma^2)$ in which $\sigma^2$ is known, the unrestricted estimator of $\mu$ is the sample mean $(\bar{Y}= \sum_{i=1}^{n} y_i /n)$.
If we know that $\mu \leq b$, the value of $\bar{Y}$ may not satisfy in that range.
 Thus, the modification of the estimator based on the prior information, $\mu \leq b$, will lead to the restricted estimator $\hat{\mu}= \bar{Y}I(\bar{Y}\leq b)+ bI(\bar{Y}>b)$.
Consequently, our prior information in \eqref{H0} is combined with the typical choice of the prior distribution and the following distribution is considered
\begin{equation}\label{TMVN}
  \pmb{\beta} \sim TN_p (\pmb{\mu}_{\pmb{\beta}}, \pmb{\Sigma}_{\pmb{\beta}}, \pmb{H}, \pmb{G}),
\end{equation}
where $TN_p (\pmb{\mu}_{\pmb{\beta}}, \pmb{\Sigma}_{\pmb{\beta}}, \pmb{H}, \pmb{G})$ denotes the truncated multivariate normal distribution with the following probability density:
\begin{equation}\label{DTVM}
  \pi(\pmb{\beta})= \dfrac{\exp \big\{(\pmb{\beta}-\pmb{\mu}_{\pmb{\beta}})^T\pmb{\Sigma}_{\pmb{\beta}}^{-1}(\pmb{\beta}-\pmb{\mu}_{\pmb{\beta}})\big\}}{\int_{\pmb{H}\pmb{\beta} \leq \pmb{G}}\exp \big\{(\pmb{\beta}-\pmb{\mu}_{\pmb{\beta}})^T\pmb{\Sigma}_{\pmb{\beta}}^{-1}(\pmb{\beta}-\pmb{\mu}_{\pmb{\beta}})\big\} d\pmb{\beta}}I(\pmb{H}\pmb{\beta} \leq \pmb{G}).
\end{equation}
By using the above prior distribution, the posterior distribution of $\pmb{\beta}$ will be:
\begin{align}\label{post}
  \pi(\pmb{\beta}|Y, X, \gamma)& =  \ell (y|\pmb{\beta} , \gamma, X) \pi(\pmb{\beta}) \nonumber \\
 & \propto \bigg[ \sum_{i=1}^{n} \bigg\{\gamma\mu_i  \log(\dfrac{y_i}{1-y_i})- \big\{ \log (\Gamma( \gamma\mu_i)) + \log \big(\Gamma\big(\gamma(1-\mu_i)\big)\big)\}\bigg\}\bigg] \nonumber \\
 & \qquad \times \exp \big\{(\pmb{\beta}-\pmb{\mu}_{\pmb{\beta}})^T\pmb{\Sigma}_{\pmb{\beta}}^{-1}(\pmb{\beta}-\pmb{\mu}_{\pmb{\beta}})\big\}I(\pmb{H}\pmb{\beta} \leq \pmb{G}).
\end{align}
 Obviously, the posterior distribution will not have a closed-form.
 Therefore, a random sample from \eqref{post} must be generated to obtain the estimator based on the loss function.
\cite{Ghosal} showed that if the likelihood function is assumed to be as follows:
\begin{equation}\label{Ghoasl:form}
  \ell(Y| \pmb{\beta}, X) \propto \exp \{Y^T X \pmb{\beta}-\sum_{i=1}^{n} \Upsilon(X_i^T \pmb{\beta}) \}
\end{equation}
in which $\Upsilon(.)$ is a positive valued convex and invertible function such that $\Upsilon''(z)>0 \quad  \forall z$, the product slice sampling method is an efficient way to obtain the sample from posterior distributions.
When the likelihood function of the Beta regression model in \eqref{log:lik} is compared to the likelihood function in  \eqref{Ghoasl:form}, it is revealed that the conditions required to employ the \cite{Ghosal}'s algorithm are not met in the Beta regression models. \\
We use the Metropolis-Hastings algorithm to derive the Beta Bayesian linear inequality restricted estimator (BBIRE).
The proposal distribution of the algorithm is specified as
\begin{equation}\label{prop:dist}
  \pmb{\beta} \sim \mathcal{TN}_p (\pmb{\beta}^{(t-1)}, \pmb{\Sigma}_{pro}, \pmb{H}, \pmb{G}).
\end{equation}
where $ \pmb{\Sigma}_{pro}$  is a $p\times p$ positive defined matrix.
Many algorithms have been proposed to generate samples from truncated multivariate normal distribution subject to linear inequality restrictions such as, \cite{Geweke1, Geweke2}, \cite{Rod}, \cite{Pakman}, \cite{Lan}, \cite{Cong}, etc., but most of them are practicable when the number of restrictions $q<p$.
The employed sampling approach in this study is based on the work of \cite{Li}. This method involves generating samples from a truncated multivariate normal distribution using a series of Gibbs cycles. Sampling from the truncated univariate normal distributions is carried out in each cycle, employing efficient customized rejection sampling techniques that are contingent upon the specific restriction type.

In the subsequent section, we compute the Beta Bayesian linear inequality restriction estimator (BBIRE) using the sample generated from the Metropolis-Hastings algorithm mentioned earlier. Our analysis reveals that the proposed estimator outperforms existing methods, even in the presence of multicollinearity concerns within the dataset.

\section{Simulation Study}\label{sec4}
In this section, the performance of the proposed estimator is illustrated by using two simulated data scenarios.
For \textbf{Scenario A}, the covariates are independent or there is a weak inter-correlation among them.
Furthermore, for \textbf{Scenario B}, a high inter-correlation is supposed for the covariates.

\subsection{Random Data Generation}
In both scenarios, the predictor function and mean are considered as follows:
\begin{align}\label{sim:mdl}
  \eta_i & = X_{i1} \beta_1+ X_{i2} \beta_2+X_{i3} \beta_3+X_{i4} \beta_4, \\
  \mu_i & = \dfrac{\exp\{\eta_i\}}{1+\exp\{\eta_i\}}; \qquad i: 1, 2, \cdots, n
\end{align}
We generate the observation of the covariates form multivariate normal distribution with mean  $\pmb{0}=(0,0,0,0)^T$ and the covariance matrix $ \pmb{\mathbb{C}}$ where $\pmb{\mathbb{C}}_{ij}= \rho^{|i-j|};\quad i, j: 1, 2, \cdots, 4$. The parameter $\rho$ controls the intensity of inter-correlation among the covariates.
The true value of the regression coefficients are chosen as $\pmb{\beta}= (1,1,1,1)^T$.
The effect of sample size on the performance of BBIRE over the other estimators is also investigated as values are taken to be $20$ and $50$.
The different values of precision parameters taken are $5$ and $10$.
Finally, the observations of the response variable are generated from $Beta(\mu_i \gamma, (1-\mu_i)\gamma)$.

It is possible that zeros and ones are observed in the generated data. To avoid these data, the recommendation of \cite{Smithson} is utilized by re-scaling the values of the dependent variable  by the following:
\begin{equation}
\tilde{Y_i}= \dfrac{Y_i(n-1)+0.5}{n}.
\end{equation}
Based on the value of parameters, the interested inequality restrictions are considered as:
\begin{align*}
\beta_1 \leq 1.5, \\
\beta_1-\beta_2+\beta_3 \leq 1.5, \\
\beta_3 \leq 1.5.
\end{align*}

\subsection{Specializing Hyperparameters}
The hyperparameters in \eqref{TMVN} are set as
\begin{equation}\label{Hyper:par}
  \pmb{\mu}_{\pmb{\beta}}= \pmb{0}\qquad \text{and} \qquad \pmb{\Sigma}_{\pmb{\beta}}= (X^T X)^{-1}
\end{equation}
and for the proposal distribution, the matrix covariance is set as the inverse of the Fisher information matrix
\begin{equation}\label{Sig:prop}
  \pmb{\Sigma}_{pro}= I^{-1}(\pmb{\beta})= \dfrac{1}{\gamma} (X^T \hat{\pmb{C}} X)^{-1}
\end{equation}
where $ \hat{\pmb{C}}$ is calculated by using the BMLE.
In addition to BBIRE and ordinary estimators used in both scenarios, the Beta Bayesian unrestricted estimator (BBUNE) is also obtained by using the multivariate normal distribution with the parameters in \eqref{Hyper:par} as prior distribution of $\pmb{\beta}$.

\subsection{Criteria for Evaluating the Estimators}
 After designing our experiment, the criteria for comparing the estimators are defined. For determining the proposed Bayesian estimator, the mean of simulated data is used here which means to suppose the squared error loss function. As the loss function is applied to find the proposed estimator, the mean squared error of the estimators obtained through 100 replicated data sets is calculated in both scenarios to display the performance of the proposed Bayesian estimator compared with the alternative estimators.
The MSE of an estimator of $\beta_j$, e.g. $\hat{\beta}_j$, is calculated as follows:
\begin{equation}
MSE(\hat{\beta}_j)= \dfrac{1}{100} \sum_{k=1}^{100} (\hat{\beta}_{kj} - \beta_j^{true})^2,
\end{equation}
where $\hat{\beta}_{kj}$ is the estimation of $\beta_j$ in the $k$th replication.
 Another criterion assumed by setting the proposed Bayesian estimator (BBIRE) as a benchmark, is the relative efficiency (RE) which is obtained by the following:
\begin{equation}\label{RE}
 RE(\hat{\beta}_j)= \dfrac{MSE(\hat{\beta}_j)}{MSE(\hat{\beta}_{j(BBIRE)})}.
\end{equation}

\subsection{Simulation Results}
This section reports the results of the simulated experiment. In each scenario, 10000 samples  are generated by the Metropolis-Hastings sampler and discounted 1000 first simulated data as burn-in to eliminate the effect of initial values. The R programming language, based on \textbf{betareg} (\cite{Cribari}) and \textbf{tmvmnorm} (\cite{Ma}) packages, is used to set up our simulation study. It is noteworthy that the convergency of the simulated Markov Chain in both scenarios has been checked by using three different initial values and the results did not reveal any problems.

\subsubsection*{Results of Scenario A}
 In this scenario, $\rho$ is set to 0 and $0.5$, respectively, with no and weak inter-correlation among the covariates.
The estimators chosen to compare the proposed Bayesian estimator with Beta Bayesian unrestricted estimator (BBUNE) and BMLE.
Table \ref{tab:table1} and \ref{tab:table2} display the estimate, standard deviation (SD), MSE and RE of each coefficient of the beta regression model based on 100 replications and different values of precision parameters. Tables show that  the proposed Bayesian estimator of each parameter outperforms other estimators in terms of SD, MSE and RE. On the other hand, when the sample size increases the SD, MSE and often RE decreases but still results show the superiority of the proposed Bayesian estimator. 

 \subsubsection*{Results of Scenario B}
The high inter-correlation considered among the covariates in this scenario is achieved by setting $\rho$ as $0.90$ and $0.95$.
 The estimators chosen to compare the proposed Bayesian estimator within this scenario, are the Beta ridge estimation (BRE) and Beta Bayesian unrestricted estimator (BBUNE).
In order to determine the ridge parameter, we examined some of the estimators that have come in the literature and finally selected the one with the lowest MSE for the designed experiment.
Suppose that $\pmb{E}$ is the matrix whose columns are the eigenvectors of $X^T \hat{\mathbf{C}} X$, $\lambda=(\lambda_1, \cdots, \lambda_p)^T $ is the eigenvalues of matrix $X^T \hat{\mathbf{C}} X$ and also $\alpha=(\alpha_1, \cdots, \alpha_p)^T=\pmb{E} \hat{\pmb{\beta}}_{BMLE}$. Thus, the chosen ridge parameter estimator is:
\begin{equation*}
  \hat{k}= \dfrac{\lambda_{max}}{\gamma \alpha_{max}^2}
\end{equation*}
in which $\alpha_{max}^2= max_j(\alpha_j^2)$ and $\lambda_{max}=max_j (\lambda_j)$.
The estimates, standard deviation (SD), MSE and RE of each coefficient of the beta regression model based on 100 replications when $\gamma=5$ are presented in Table \ref{tab:table3} and when $\gamma=10$ are presented in Tables \ref{tab:table4}.
 The results show that BBIRE produces much lower MSE and SD of the estimates for all coefficients compared to BRE and BBUNE.
 As it is obvious, the SD, MSE and RE of each coefficient decrease when the sample size increases.
The results show the incorporation of linear inequality restrictions in Bayesian inference results in sparsity without using any regularization.

\section{Application to Real Data} \label{sec5}
The utilization of the suggested technique is exemplified by a research conducted on the well-being index of Turkey in 2015, as documented by Aktas (year). The index encompasses several dimensions, including housing employment, income and wealth, health, education, climate, protection, public involvement, and access to community assets and social life. The life satisfaction index ranges from zero to one.
Values that approach unity are indicative of a higher quality of life. The data was sourced from the official website of the Turkish Statistical Institute.

We are interested in nine indicators from 41 indicators within the data set as mentioned by \cite{Abonazel}. The chosen indicators as the covariates of the model are
number of rooms per person $(X_1)$,
percentage of households declaring to fail on meeting basic needs $(X_2)$,
satisfaction rate with public health services$(X_3)$,
average point of necessary placement scores of the system for transition to secondary education from basic education $(X_4)$,
satisfaction rate with public education services $(X_5)$,
percentage of the population receiving waste services $(X_6)$,
satisfaction rate with public safety services $(X_7)$,
the access rate of the population to sewerage and pipe system $(X_8)$,
and finally, we consider the level of happiness as the response variable.

First, the intensity of correlation among the covariates is investigated. Table \ref{Tab:cor} reports the correlation matrix of covariates. There is
 strong correlation between some covariates. Therefore, the proposed Bayesian restricted estimator, the Beta Bayesian unrestricted estimator (BBUNE), the Beta maximum likelihood estimator (BMLE) and the Beta ridge estimator (BRE) are calculated.
 
 The restrictions on the model parameters based on the conclusions of the researches done by \cite{Aktas} and \cite{Abonazel} are selected as
\begin{equation}
  \beta_2 \leq0, \beta_3\geq 0, \beta_5\geq 0, \beta_6\leq0.
\end{equation}

Similarity to Section \ref{sec4}, the number of samples generated by the algorithm described in Section \ref{sec3} is 10000 data with discounting 1000 first data as burn-in and also the parameters of the prior distribution and proposed distribution are chosen as Section \ref{sec4}. For the ridge parameter, due to results presented in \cite{Abonazel}, the following is considered
 \begin{equation*}
  \hat{k}= \dfrac{\lambda_{min}}{\gamma \alpha_{min}^2}
\end{equation*}
 in which $\alpha_{min}^2= min_j(\alpha_j^2)$ and $\lambda_{min}=min_j (\lambda_j)$.
 In order to compare the performance of the proposed Bayesian restricted estimator, the bootstrap case re-sampling method is used.
A bootstrap sample size of 30 from the 81 observations of the data set with 100 replacements is chosen. The estimators, standard deviations, and relative efficiencies according to each bootstrap sample are computed.
  The estimators and the standard deviations are estimated using the sample mean and sample standard deviation.

\begin{landscape}
\begin{table}
  \begin{center}
    \caption{Results of simulation for Scenario A when $\gamma=5$.}
    \label{tab:table1}
 \footnotesize
\begin{tabular}{cclccccccccc}
    \toprule
            &  & &\multicolumn{4}{c}{$ n=20 $}&&\multicolumn{4}{c}{$ n=50 $}\\
            \cmidrule{4-7} \cmidrule{9-12}
 $\rho$ &  \text{Parameters} & \text{Estimators} &\text{Estimates}&\text{SD} & \text{MSE} & \text{R.E}& &\text{Estimates}&\text{SD} & \text{MSE} & \text{R.E}\\
   \hline
  \multirow{12}{*}{0}&$\beta_1$   & BMLE   &1.0329 &0.2397 &0.0580 &2.1993 &&1.0065 &0.1552 &0.0239 &1.8570    \\
                                 &                 & BBUNE &1.0002 &0.2195 &0.0477 &1.8098 &&0.9909 &0.1387 &0.0191 &1.4880    \\
                                 &                 & BBIRE  &0.9341 &0.1491 &0.0264 &-          &&0.9780 &0.1118 &0.0129 &-    \\[5pt]
                                 &$\beta_2$  & BMLE   &1.0465 &0.2521 &0.0651 &2.2703 &&1.0305 &0.1395 &0.0202 &1.5640    \\
                                 &                 & BBUNE &1.0151 &0.2245 &0.0501 &1.7484 &&1.0138 &0.1291 &0.0167 &1.2923   \\
                                 &                 & BBIRE  &1.0438 &0.1644 &0.0287 &-          &&1.0194 &0.1125 &0.0129 &-   \\[5pt]
                                 &$\beta_3$  & BMLE   &1.0708 &0.2320 &0.0583 &2.9931 &&0.9790 &0.1483 &0.0222 &1.5158   \\
                                 &                 & BBUNE &1.0335 &0.2002 &0.0408 &2.0938 && 0.9625 &0.1405 &0.0210 &1.4296 \\
                                 &                 & BBIRE  &0.9578 &0.1337 &0.0195 &-          &&0.9589 &0.1144 &0.0147 &-  \\[5pt]
                                 &$\beta_4$  & BMLE   &1.0141 &0.2642 &0.0693 &2.0318 &&1.0263 &0.1593 &0.0258 &1.9179  \\
                                 &                 & BBUNE &0.9839 &0.2359 &0.0554 &1.6231 &&1.0082 &0.1429 &0.0203 &1.5075   \\
                                 &                 & BBIRE  &0.9818 &0.1847 &0.0341 &-          &&1.0080 &0.1163 &0.0135 &-   \\[5pt]
                                 \hline
 \multirow{12}{*}{0.5}&$\beta_1$ & BMLE   &0.9497 &0.2831 &0.0819 &1.6374 &&0.9905 &0.1494 &0.0222 &1.7998\\
                                 &                 & BBUNE &0.9173 &0.2732 &0.0807 &1.6137 &&0.9722 &0.1332 &0.0183&1.4861\\
                                 &                 & BBIRE  &0.8693 &0.1824 &0.0500 &   -       &&0.9621 &0.1049 &0.0123&-\\[5pt]
                                 &$\beta_2$  & BMLE   &0.9559 &0.3076 &0.0956 &2.5631 &&0.9935 &0.1510 &0.0226 &1.8423\\
                                 &                 & BBUNE &0.9323 &0.2782 &0.0812 &2.1765 &&0.9753 &0.1363 &0.0190&1.5467\\
                                 &                 & BBIRE  &1.0072 &0.1940 &0.0373 &    -      && 0.9948 &0.1113 &0.0123&-\\[5pt]
                                 &$\beta_3$ & BMLE    &1.0107 &0.3393 &0.1141 &2.4199 &&0.9623 &0.1498 &0.0236 &1.5364\\
                                 &                 & BBUNE &0.9707 &0.2921 &0.0853 &1.8099 &&0.9453 &0.1385 &00.0220&1.4301\\
                                 &                 & BBIRE  &0.8989 &0.1932 &0.0472 &  -        &&0.9346 &0.1059 &0.0154&- \\[5pt]
                                 &$\beta_4$  & BMLE   &0.9605 &0.3487 &0.1220 &2.0149 &&0.9574 &0.1594 &0.0270 &1.5837\\
                                 &                 & BBUNE &0.9279 &0.2976 &0.0929 &1.5346 && 0.9386 &0.1447 &0.0245&1.4394 \\
                                 &                 & BBIRE  &0.9574 &0.2435 &0.0605 &  -        &&0.9482 &0.1204 &0.0170&-\\[5pt]
  \hline
\end{tabular}
\end{center}
\end{table}
\end{landscape}

\begin{landscape}
\begin{table}
  \begin{center}
    \caption{Results of simulation for Scenario A when $\gamma=10$.}
    \label{tab:table2}
 \footnotesize
\begin{tabular}{cclccccccccc}
    \toprule
            &  & &\multicolumn{4}{c}{$ n=20 $}&&\multicolumn{4}{c}{$ n=50 $}\\
            \cmidrule{4-7} \cmidrule{9-12}
 $\rho$ &  \text{Parameters} & \text{Estimators} &\text{Estimates}&\text{SD} & \text{MSE} & \text{R.E}& &\text{Estimates}&\text{SD} & \text{MSE} & \text{R.E}\\
   \hline
  \multirow{12}{*}{0}&$\beta_1$ & BMLE     &1.0210 &0.1975 &0.0391 &2.3230  &&0.9850 &0.1248 &0.0157 &1.6030   \\
                                 &                 & BBUNE & 1.0068 &0.1733 &0.0298 &1.7701 &&0.9862 &0.1153 &0.0134 &1.3676   \\
                                 &                 & BBIRE  & 0.9681 &0.1263 &0.0168 &-          &&0.9760 &0.0963 &0.0098 &-   \\[5pt]
                                 &$\beta_2$  & BMLE   & 1.0570 &0.2302 &0.0557 &2.3686 &&0.9937 &0.1192 &0.0141 &1.5781   \\
                                 &                 & BBUNE & 1.0404 &0.2074 &0.0442 &1.8806 &&0.9946 &0.1040 &0.0107 &1.2020   \\
                                 &                 & BBIRE  & 1.0450 &0.1473 &0.0235 &-          &&0.9995 &0.0950 &0.0089 &-   \\[5pt]
                                 &$\beta_3$ & BMLE    & 1.0503 &0.2264 &0.0533 &2.8775 &&1.0083 &0.1152 &0.0132 &1.5597   \\
                                 &                 & BBUNE & 1.0361 &0.2072 &0.0438 &2.3661 &&1.0099 &0.1066 &0.0114 &1.3401   \\
                                 &                 & BBIRE  & 0.9844 &0.1358 &0.0185 &-          &&1.0012 &0.0925 &0.0085 &-   \\ [5pt]
                                 &$\beta_4$  & BMLE   & 1.0182 &0.2097 &0.0439 &1.9941 &&0.9794 &0.1301 &0.0172 &1.4514    \\
                                 &                 & BBUNE & 1.0053 &0.1898 &0.0357 &1.6232 &&0.9807 &0.1182 &0.0142 &1.1997   \\
                                 &                 & BBIRE  & 0.9978 &0.1490 &0.0220 &-          &&0.9808 &0.1076 &0.0118 &-   \\ [5pt]
                                 \hline
 \multirow{12}{*}{0.5}&$\beta_1$ & BMLE   &1.0161 &0.2483 &0.0613 &2.3705   &&0.9894 &0.1335 &0.0177 &1.6373   \\
                                 &                 & BBUNE &0.9906 &0.2297 &0.0523 &2.0222   &&0.9854 &0.1267 &0.0161 &1.4849   \\
                                 &                 & BBIRE  &0.9480 &0.1530 &0.0259 &-            &&0.9746 &0.1015 &0.0108 &-   \\[5pt]
                                 &$\beta_2$  & BMLE   &1.0313 &0.2565 &0.0661 &2.2718   && 1.0023 &0.1361 &0.0183 &1.6052   \\
                                 &                 & BBUNE &1.0015 &0.2412 &0.0576 &1.9785   &&0.9962 &0.1262 &0.0158 &1.3803   \\
                                 &                 & BBIRE  &1.0370 &0.1674 &0.0291 &-            &&1.0078 &0.1071 &0.0114 &-  \\[5pt]
                                 &$\beta_3$ & BMLE    &1.0056 &0.2364 &0.0554 &1.9416   &&1.0075 &0.1313 &0.0171 &1.4638   \\
                                 &                 & BBUNE &0.9780 &0.2304 &0.0530 &1.8593   &&1.0039 &0.1276 &0.0161 &1.3786   \\
                                 &                 & BBIRE  &0.9282 &0.1536 &0.0285 &-            &&0.9899 &0.1082 &0.0117 &-   \\ [5pt]
                                 &$\beta_4$  & BMLE   &1.0459 &0.2366 &0.0575 &2.0748   && 0.9829 &0.1397 &0.0196 & 1.6578  \\
                                 &                 & BBUNE &1.0193 &0.2048 &0.0419 &1.5121   &&0.9768 &0.1258 &0.0162 &1.3691   \\
                                 &                 & BBIRE  &1.0185 &0.1663 &0.0277 &-            &&0.9771 &0.1069 &0.0118 &-\\[5pt]
  \hline
\end{tabular}
\end{center}
\end{table}
\end{landscape}

\begin{landscape}
\begin{table}
  \begin{center}
    \caption{Results of simulation for Scenario B when $\gamma=5$.}
    \label{tab:table3}
 \footnotesize
\begin{tabular}{cclccccccccc}
    \toprule
            &  & &\multicolumn{4}{c}{$ n=20 $}&&\multicolumn{4}{c}{$ n=50 $}\\
            \cmidrule{4-7} \cmidrule{9-12}
 $\rho$ &  \text{Parameters} & \text{Estimators} &\text{Estimates}&\text{SD} & \text{MSE} & \text{R.E}& &\text{Estimates}&\text{SD} & \text{MSE} & \text{R.E}\\
   \hline
  \multirow{12}{*}{0.9} &$\beta_1$ &  BRE   &0.9341 &0.3719 &0.1412 &1.7573 && 0.8346 &0.2583 &0.0934 &1.4753\\
                                 &                 & BBUNE  &1.0133 &0.4611 &0.2107 & 2.6210&& 0.8656 &0.2926 &0.1029 &1.6244\\
                                 &                 & BBIRE  &0.8663 &0.2513 &0.0804 &-&& 0.8338 &0.1899 &0.0633 &-\\[5pt]
                                 &$\beta_2$  &  BRE     &0.8310 &0.4477 &0.2270 &1.8872&& 0.8897 &0.2695 &0.0841 &2.2456\\
                                 &                 & BBUNE &0.8792 &0.5434 &0.3070 & 2.5520&& 0.9230 &0.2962 &0.0928 &2.4779\\
                                 &                 & BBIRE  & 1.0699 &0.3414 &0.1203 &-&& 1.0237 &0.1930 &0.0374 &-\\[5pt]
                                 &$\beta_3$  & BRE     &0.8569 &0.4190 &0.1943 &1.2818&& 0.8919 &0.2758 &0.0870 &1.4902\\
                                 &                 & BBUNE &0.9292 &0.5336 &0.2869 & 1.8930 && 0.9306 &0.3126 &0.1016 &1.7393\\
                                 &                 & BBIRE  &0.7829 &0.3248 &0.1516 &-&& 0.8721 &0.2060 &0.0584 &-\\[5pt]
                                 &$\beta_4$  & BRE     &0.8042 &0.4148 &0.2087 &1.1899 &&0.9113 &0.2490 &0.0692 &1.2982 \\
                                 &                 & BBUNE &0.8534 &0.5265 &0.2959 & 1.6872&&  0.9542 &0.2886 &0.0846 &1.5853 \\
                                 &                 & BBIRE  &0.9483 &0.4177 &0.1754 &-&& 0.9928 &0.2320 &0.0533 &-\\[5pt]
                                 \hline
\multirow{12}{*}{0.95}&$\beta_1$ &  BRE    &1.1185 &0.3966 &0.1697 &4.1024  &&0.8015 &0.3506 &0.1611 &1.2388\\
                                 &                 & BBUNE & 1.2658 &0.5440 &0.3636 &8.7876&& 0.8181 &0.4249 &0.2118 &1.6294\\
                                 &                 & BBIRE  & 0.9148 &0.1856 &0.0414 &-&&  0.7721 &0.2808 &0.1300 &- \\[5pt]
                                 &$\beta_2$  & BRE     & 0.8060 &0.5961 &0.3894 &1.1995 &&0.8990 &0.3198 &0.1114 &1.5408\\
                                 &                 & BBUNE &0.8691 &0.7998 &0.6504 & 2.0033&& 0.9324 &0.3970 &0.1606 & 2.2204 \\
                                 &                 & BBIRE  &1.2061 &0.5339 &0.3247 &- && 1.0523 &0.2651 &0.0723 &-\\[5pt]
                                 &$\beta_3$ &  BRE     & 0.7800 &0.5615 &0.3606 &1.0633 &&0.9015 &0.4007 &0.1686 &1.5189\\
                                 &                 & BBUNE &0.7741 &0.7847 &0.6606 &1.9481 && 0.9408 &0.4757 &0.2275 &2.0492  \\
                                 &                 & BBIRE  &0.6047 &0.4297 &0.3391 &-&& 0.8488 &0.2984 &0.1110 &- \\[5pt]
                                 &$\beta_4$  & BRE     & 0.7568 &0.5782 &0.3902 &0.8297 &&0.8865 &0.4017 &0.1726 &1.0589\\
                                 &                 & BBUNE &0.7276 &0.8045 &0.7149 &1.5202&& 0.9370 &0.4806 &0.2326 &1.4270 \\
                                 &                 & BBIRE  &0.9049 &0.6826 &00.4703 &-&& 0.9759 &0.4051 &0.1630 &-\\[5pt]
  \hline
\end{tabular}
\end{center}
\end{table}
\end{landscape}

\begin{landscape}
\begin{table}
  \begin{center}
    \caption{Results of simulation for Scenario B when $\gamma=10$.}
    \label{tab:table4}
 \footnotesize
\begin{tabular}{cclccccccccc}
    \toprule
            &  & &\multicolumn{4}{c}{$ n=20 $}&&\multicolumn{4}{c}{$ n=50 $}\\
            \cmidrule{4-7} \cmidrule{9-12}
 $\rho$ &  \text{Parameters} & \text{Estimators} &\text{Estimates}&\text{SD} & \text{MSE} & \text{R.E}& &\text{Estimates}&\text{SD} & \text{MSE} & \text{R.E}\\
   \hline
  \multirow{12}{*}{0.90}&$\beta_1$ & BRE     &0.9311 &0.3373 &0.1174 &1.5126 && 0.9757 &0.2252 &0.0508 &2.0275 \\
                                 &                 & BBUNE &0.9664 &0.4061 &0.1644 &2.1192 && 0.9891 &0.2341 &0.0544 &2.1708\\
                                 &                 & BBIRE  &0.8187 &0.2125 &0.0776 &-          && 0.9339 &0.1445 &0.0251 & -\\[5pt]
                                 &$\beta_2$  &  BRE     &0.8678 &0.4338 &0.2038 &1.8568 &&0.8907 &0.2438 &0.0708 &2.8426\\
                                 &                 & BBUNE &0.8867 &0.5340 &0.2951 &2.6889 && 0.8991 &0.2571 &0.0756 &3.0377\\
                                 &                 & BBIRE  &1.1118 &0.3134 &0.1097 &-          &&1.0066 &0.1585 &0.0249 & -\\[5pt]
                                 &$\beta_3$  & BRE     &0.9305 &0.4341 &0.1914 &1.7062 &&0.9480 &0.2630 &0.0712 &1.7666\\
                                 &                 & BBUNE &0.9590 &0.5072 &0.2563 &2.2851 &&0.9589 &0.2754 &0.0768 &1.9054\\
                                 &                 & BBIRE  &0.7894 &0.2617 &0.1122 &-          &&0.8984 &0.1740 &0.0403 & -\\[5pt]
                                 &$\beta_4$  &  BRE     &0.9554 &0.4040 &0.1636 &1.0711  &&0.9351 &0.2444 &0.0634 &1.4894\\
                                 &                 & BBUNE &0.9901 &0.4844 &0.2324 &1.5216  &&0.9469 &0.2553 &0.0674 &1.5836 \\
                                 &                 & BBIRE  &1.0993 &0.3799 &0.1527 &-           &&0.9876 &0.2069 &0.0425 & -\\[5pt]
                                 \hline
\multirow{12}{*}{0.95}&$\beta_1$ &  BRE      & 1.1144 &0.4113 &0.1806 &3.1281  &&0.9480 &0.3256 &0.1077 & 1.6747\\
                                 &                 & BBUNE & 1.1903 &0.5195 &0.3034 &5.2563  && 0.9709 &0.3587 &0.1283 &1.9948\\
                                 &                 & BBIRE  &  0.9177 &0.2269 &0.0577 &- &&0.8743 &0.2213 &0.0643 &-\\[5pt]
                                 &$\beta_2$  &  BRE     & 0.7790 &0.5356 &0.3329 &1.8111  &&0.9026 &0.2972 &0.0969 & 2.2706 \\
                                 &                 & BBUNE &  0.7389 &0.6536 &0.4911 &2.6722 && 0.9234 &0.3237 &0.1096 &2.5666 \\
                                 &                 & BBIRE  & 1.0844 &0.4224 &0.1838 &-   && 1.0814 &0.1909 &0.0427 &- \\[5pt]
                                 &$\beta_3$ &  BRE     &  0.8875 &0.4474 &0.2108 &1.1278 &&0.9398 &0.3477  &0.1233 & 1.7289\\
                                 &                 & BBUNE &  0.8935 &0.5782 &0.3423 &1.8312  && 0.9603 &0.3651  &0.1336 &1.8726 \\
                                 &                 & BBIRE  &  0.7380 &0.3457 &0.1869 &-  && 0.8610 &0.2292 &0.0713 &- \\[5pt]
                                 &$\beta_4$  &  BRE     &  0.8909 &0.5385 &0.2990 &0.9037 &&0.8976 &0.3346  &0.1213 & 1.5707\\
                                 &                 & BBUNE &  0.9388 &0.6701 &0.4483 &1.3550  && 0.9057 &0.3547  &0.1334 &1.7278 \\
                                 &                 & BBIRE  &  1.0400 &0.5767 &0.3308 &-  && 0.9809 &0.2787 &0.0772 &-\\[5pt]
  \hline
\end{tabular}
\end{center}
\end{table}
\end{landscape}

\begin{table}
  \begin{center}
    \caption{Correlation matrix of covariates of real data set.}
    \label{Tab:cor}
 \footnotesize
\begin{tabular}{c|cccccccc}
    \toprule
 & $X_1$ & $X_2$ & $X_3$ & $X_4$ & $X_5$ & $X_6$ & $X_7$ & $X_8$ \\
     \hline
$X_1$ &1.000 &-0.822  &0.572  &0.877  &0.416  &0.154  &0.480  &0.197\\
$X_2$ &      &  1.000 &-0.585 &-0.759 &-0.372 &-0.182 &-0.399 &-0.248\\
$X_3$ &      &        & 1.000 & 0.497 & 0.839 & 0.039 & 0.843 & 0.110 \\
$X_4$ &      &        &       &  1.000 & 0.312 & 0.174 & 0.414 & 0.241\\
$X_5$ & & &  &  &  1.000 &-0.173 & 0.891& -0.112\\
$X_6$ & & &  &  & &  1.000& -0.225&  0.931\\
$X_7$ & & &  &  &  & &  1.000 &-0.159\\
$X_8$ & & &  &  & &  & &  1.000\\
    \hline
\end{tabular}
\end{center}
\end{table}

   Summary statistics for all parameters are presented in Table \ref{real1}. The results of the tables illustrate that the Beta Bayesian inequality restricted estimator has the lowest standard  deviation. Due to comparing the performance of the proposed Bayesian estimator, Table \ref{real2} presents the total simulated relative efficiency which is calculated by the following formula:

 \begin{equation}\label{TSRE}
   TSRE(\hat{\pmb{\beta}})= \dfrac{\sum_{j=0}^{8} MSE(\hat{\beta_j})}{\sum_{j=0}^{8} MSE(\hat{\beta}_{j(BBIRE)})}
 \end{equation}
Since the value of TSRE for all traditional estimators are larger than one, it indicates that BBIRE outperforms the other estimators.
 \vspace{10pt}
\begin{table}
  \begin{center}
    \caption{Bootstrapped estimates and standard deviation (SD) of model parameters for real data set.}
    \label{real1}
 \footnotesize
\begin{tabular}{ccccccccccccc}
    \toprule
  &\multicolumn{2}{c}{\text{BMLE}}&&\multicolumn{2}{c}{\text{BRE}}&&\multicolumn{2}{c}{\text{BBUNE}}&&\multicolumn{2}{c}{\text{BBIRE}}\\
            \cmidrule{2-3} \cmidrule{5-6}\cmidrule{8-9} \cmidrule{11-12}
  &\text{Estimates}&\text{SD} &&\text{Estimates}&\text{SD} &&\text{Estimates}&\text{SD} &&\text{Estimates}&\text{SD}\\
   \hline
 intercept &2.7208 &2.2456  && 2.1614 &1.9728  &&  2.7598 &2.2714  &&2.9719  &1.8160\\
    $X_1$ & -0.0413&0.6825  && 0.0258 &0.6206  && -0.0404 &0.6885  &&-0.0545 &0.6248\\
   $X_2$ & -2.0861&1.2015  && -1.7173&1.1993  && -2.1067 &1.2161  &&-2.0456 &0.8096\\
   $X_3$ & 1.7496 &1.9315  && 1.4798 &1.5576  && 1.7415  &1.9315  &&2.1923  &1.1235\\
   $X_4$ & -0.6216&0.4777  && -0.5611&0.4502  && -0.6290 &0.4809  &&-0.6051 &0.4668 \\
   $X_5$ & 0.9944 &1.6243  && 0.7943 &1.3594  && 1.0076  &1.6168  &&1.6395  &0.7606\\
   $X_6$ & -0.8395&1.0613  && -0.6617&0.8476  && -0.8456 &1.0706  &&-1.2225 &0.7236\\
   $X_7$ & -1.2312&2.3902  &&-0.7414 &1.9481  && -1.2393 &2.4084  &&-2.4460 &1.7759\\
   $X_8$ & 0.3845 &1.2416  && 0.2677 &1.0157  && 0.3878  &1.2487  &&0.6641  &0.9028\\
    \hline
\end{tabular}
\end{center}
\end{table}

\begin{table}
  \begin{center}
    \caption{TSRE of estimators for real data set.}
    \label{real2}
 \footnotesize
\begin{tabular}{ccccccc}
    \toprule
     &&\text{BMLE}&&\text{BRE}&&\text{BBUNE}\\
     \hline
  \text{TSRE} && 1.8657 && 1.4719 &&1.8813 \\
    \hline
\end{tabular}
\end{center}
\end{table}

\section{Conclusions}\label{sec6}
This paper has considered the problem of Bayesian estimating parameters restricted by some linear inequality restrictions in Beta regression models.
Beta regression models with logistic link function do not satisfy the conditions of the estimation parameters method mentioned by \cite{Ghosal}. Thus,
a new method of estimating restricted parameters in the Beta regression model has been presented and is practicable for any other members of GLM as well.
The simulation results illustrated that the proposed method provides a parameter estimation that outperforms well-known estimators even if the design matrix is ill-conditioned. The real data application also shows the practicality of the proposed method in estimating the parameters.


\end{document}